\documentclass[prd, twocolumn, tightenlines, twoside, secnumarabic, longbibliography,
superscriptaddress, preprintnumbers, nofootinbib, notitlepage, fleqn]{revtex4-1}

\usepackage{hyperref}
\usepackage{amsmath,amssymb}
\usepackage{cleveref}
\usepackage{graphicx}
\usepackage{pstricks}

\newcommand{\iso}[2]{{\ensuremath{{}^{#2}}\ensuremath{\rm #1}}}

\begin{document}

\preprint{MITP/16-058}

\title{Antineutrino monitoring of spent nuclear fuel}

\author{Vedran Brdar}
\email{vbrdar@uni-mainz.de}
\affiliation{PRISMA Cluster of Excellence \& Mainz Institute for Theoretical Physics,
             Johannes Gutenberg University, Staudingerweg 7, 55099 Mainz, Germany}

\author{Patrick Huber}
\email{pahuber@vt.edu}
\affiliation{Center for Neutrino Physics, Virginia Tech, Blacksburg 24061, USA}

\author{Joachim Kopp}
\email{jkopp@uni-mainz.de}
\affiliation{PRISMA Cluster of Excellence \& Mainz Institute for Theoretical Physics,
             Johannes Gutenberg University, Staudingerweg 7, 55099 Mainz, Germany}

\date{\today}

\begin{abstract}
  Military and civilian applications of nuclear energy have left a
  significant amount of spent nuclear fuel over the past 70 years.
  Currently, in many countries world wide, the use of nuclear energy
  is on the rise. Therefore, the management of highly radioactive
  nuclear waste is a pressing issue.  In this
  letter, we explore antineutrino detectors as a tool for monitoring and
  safeguarding nuclear waste material.  We compute the flux and
  spectrum of antineutrinos emitted by spent nuclear fuel elements as a
  function of time, and we illustrate the usefulness of antineutrino
  detectors in several benchmark scenarios. In particular, we
  demonstrate how a measurement of the antineutrino flux can help to
  re-verify the contents of a dry storage cask in case the monitoring
  chain by conventional means gets disrupted. 
  We then comment on the usefulness of antineutrino detectors at
  long-term storage facilities such as Yucca mountain.
  Finally, we put forward
  antineutrino detection as a tool in locating underground ``hot spots''
  in contaminated areas such as the Hanford site in Washington state.
\end{abstract}

\maketitle

\section{Introduction}

With carbon dioxide induced climate change and the scarceness of
fossil fuels becoming imminent problems for humanity, nuclear energy
is undergoing a renaissance. However, nuclear
technology comes with a number of intrinsic problems, such as the
limited availability of nuclear fuel, the danger of proliferation of
nuclear weapons technology,
the risk of major accidents, and the management of highly radioactive
waste. As a result, nuclear energy is relatively expensive
compared to many other energy sources.

In this letter, we will in particular focus on the waste issue: we
will argue that a measurement of the antineutrino flux emitted by
beta-decaying isotopes can be a unique component in a multi-faceted
approach to monitoring and safeguarding nuclear waste repositories.
The unique advantage of antineutrinos is that they penetrate the
shielding surrounding the repository and thus offer a direct method
for remotely probing the nuclear material inside. Other probes like
gamma rays or neutrons, see for instance Ref.~\cite{Ziock}, are heavily
attenuated by the materials they need to traverse on the way to a
detector\footnote{For a recent study on muon radiography applied to the problem at hand, see~\cite{Poulson:2016fre}.}.  Therefore, relating their measured fluxes to the actual
content of the repository requires a sophisticated propagation model,
which in turn relies on an accurate knowledge of the contents of the
repository. This cyclic dependence on information is one of the major
limitations of conventional monitoring methods. On the downside, the
very fact that antineutrinos are not attenuated even by a whole
mountain, implies that antineutrino detection has to deal with very
small cross sections $\lesssim
10^{-41}$~cm$^2$~\cite{Vogel:1999zy}. Any meaningful flux measurement
thus requires the deployment of a large detector with at least several
tons of active material for a time period of order months.

Nevertheless, thanks to advances in detector technology, this appears
feasible at a comparatively reasonable cost.  In fact, practical
applications of antineutrino detectors in the nuclear industry
have been discussed for a long time, mostly in the context of
monitoring power reactors~\cite{Mikaelian:1978, Bernstein:2001cz,
Nieto:2003wd, Huber:2004xh, Bernstein:2010, Christensen:2013eza,
Christensen:2014pva}. Several detectors
have been built to demonstrate the feasibility of such
applications~\cite{Klimov:1994a, Bowden:2006hu}, and further studies
are planned in current and future experiments~\cite{Bowden:2016ntq}.

In the following, we will first compute the antineutrino flux and
spectrum emitted by spent nuclear fuel, and then consider several
scenarios in which antineutrino detectors can be used in the context
of radioactive waste repositories.

\section{Antineutrino emission from spent nuclear fuel}

\begin{figure}
  \begin{center}
    \includegraphics[width=0.8\columnwidth]{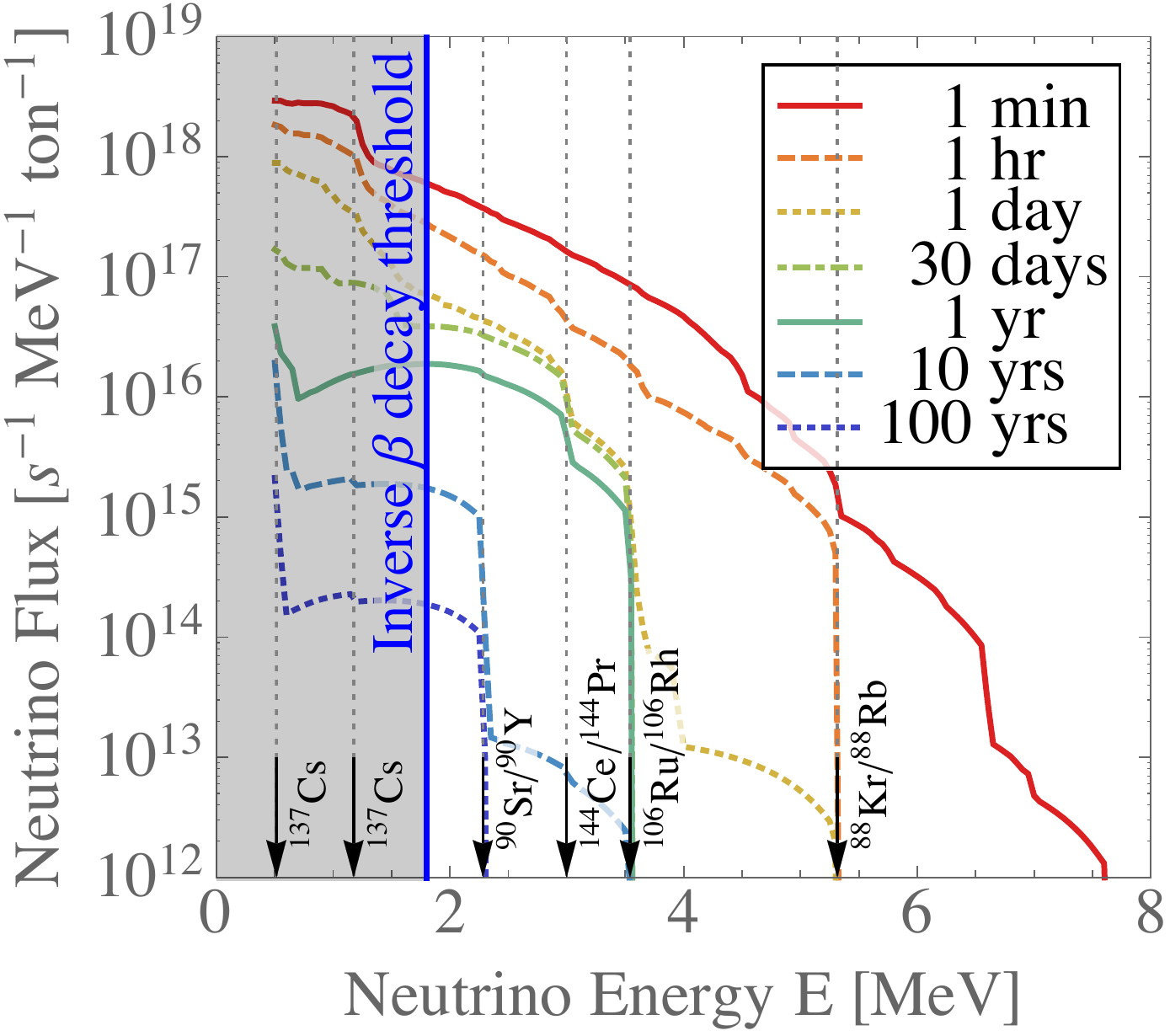}
  \end{center}
  \caption{The spectrum of electron antineutrinos emitted by spent
    nuclear fuel as a function of the time after discharge from the
    reactor. We also indicate in gray the area below the threshold for
    inverse beta decay, the dominant antineutrino detection process, at
    1.8~MeV. The data underlying this plot is available in the
    supplemental material~\cite{Huber:2016supp}.}
  \label{fig:spectra}
\end{figure}

For the first 1,000--10,000~years after discharge from a reactor,
the total activity of spent nuclear fuel is nearly exclusively caused
by beta decays (and the associated gamma emission). Therefore, a large number of
antineutrinos is produced. However, detection by inverse beta
decay, $\bar\nu_e + p \to n + e^+$, the main detection
reaction for electron antineutrinos, requires antineutrino energies
of at least 1.8\,MeV. The lifetime of a beta decaying
nucleus scales roughly like $Q^5$, where $Q$ is the energy released in
the decay. Therefore, the detectable antineutrino signal for most
fission fragments decays within hours to days after fission
ends. There is, however, a handful of isotopes that have a two stage
decay, where the first decay has very small $Q$ and thus a resulting
long lifetime, followed by a fast decay with $Q>1.8\,\text{MeV}$.
The most notable example is
strontium-90, which decays with a half-life of 28.90\,yrs to
yttrium-90, which in turn decays within hours to the stable
zirconium-90 with $Q=2.22801$\,MeV~\cite{Browne:1997cbp}. Strontium-90 is produced in
around 5\% of all fission
events~\cite{Koning:2006, SCALE, ORIGEN}. The isotopes with the next longest
lifetimes with antineutrino emission above 1.8\,MeV in their decay chains are
ruthenium-106 (371.8\,days~\cite{DeFrenne:2008kmy}) and cerium-144
(284.91\,days~\cite{NuDat}). As a result, the
detectable antineutrino emission of spent nuclear fuel after more than
a few years is entirely given by strontium-90. It is worth noting
that strontium-90 (like all other fission fragments) remains in the
high-level waste resulting from reprocessing using the widely employed
PUREX process. In \cref{fig:spectra}, we plot the number of electron
antineutrinos emitted per second, per MeV, and per ton of spent nuclear
fuel as a function of antineutrino energy for fuel elements of different
age.  We assume a
burnup\footnote{Burnup is a measure of how much energy per unit mass
  has been extracted from nuclear fuel. It is directly proportional to 
  the total number of fissions and thus to the
  strontium-90 content and the antineutrino emission rate.}
of 45~GW\,days.  As
expected, we observe a softening of the spectrum over time, as
short-lived isotopes with large $Q$ values decay away.
Note, however, that even after 100~yrs, a
non-zero flux remains above the energy threshold of 1.8~MeV for
inverse beta decay. 

\section{Dry cask storage facilities}

\begin{figure}
  \centering
  \includegraphics[width=0.95\columnwidth,clip,trim=0 55 35 0]{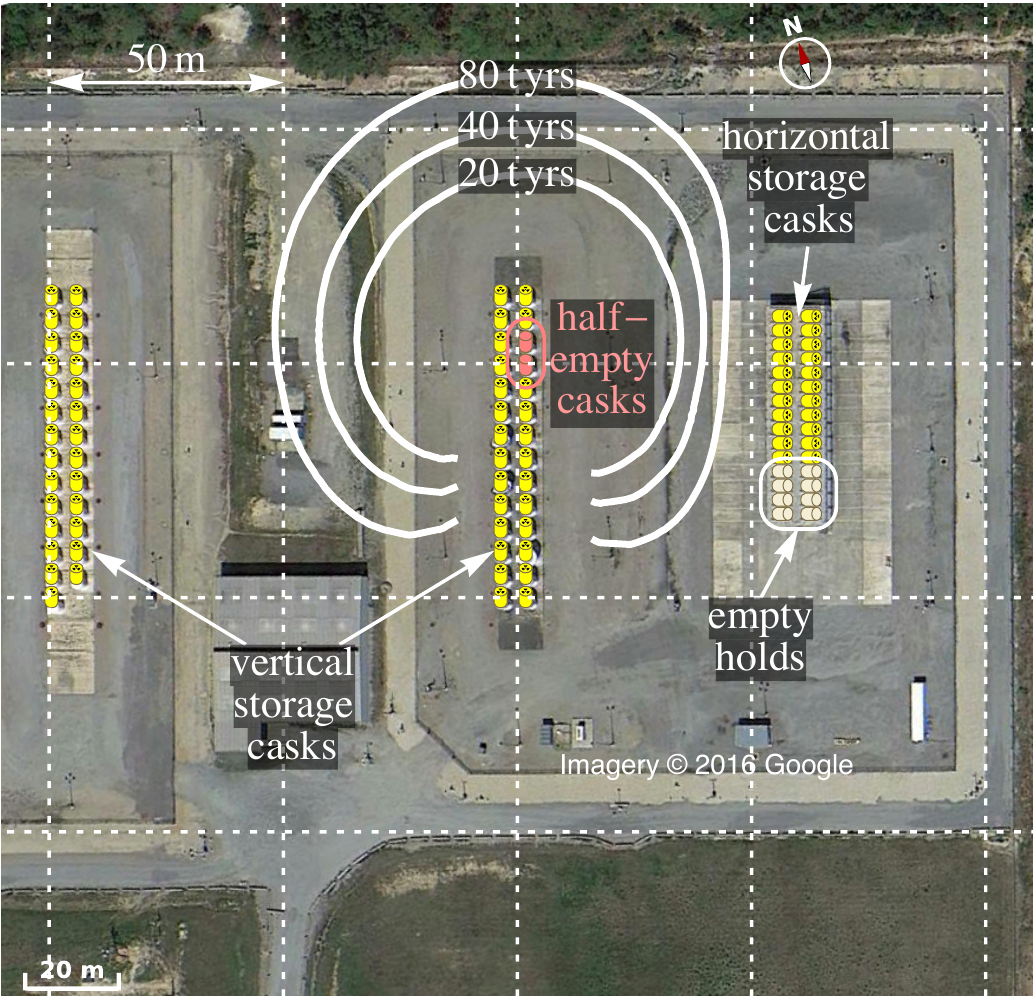}
  \caption{The dry storage facility at the Surry Nuclear Power Plant
    in Virginia, USA \cite{Greene:2013}.  Filled storage casks,
    highlighted in yellow, contain 9--16~MTU each.  In
    the benchmark scenario discussed in the text, we assume that 50\%
    of the spent fuel in two 15~MTU casks (marked in red) have gone missing.
    Colored contours indicate the exposure (in ton~yrs) required to
    establish the loss of nuclear material at the 90\% confidence
    level.}
  \label{fig:surry}
\end{figure}

As long term storage facilities
for spent nuclear fuel are becoming available only slowly, temporary
storage solutions have become a necessity.  Once fuel elements have
been allowed to cool in a spent fuel pool for $\sim 10$~yrs
\cite{Alvarez:2011,NRCweb} after discharge from the reactor, they are
typically transferred to dry storage casks, large shielded steel
cylinders several meters tall, each of them holding $\sim
14$--$24$~tons of spent nuclear fuel elements with a uranium content
of 10--17~tons \cite{Greene:2013, IDB1996, NRCweb}. The layout of a
typical dry storage facility is shown in \cref{fig:surry}.  Even
though safety and security measures are in place to protect such
facilities, manipulations are imaginable. The core of the IAEA's
(International Atomic Energy Agency's) methodology for spent fuel is
so-called continuity of knowledge (CoK): the amount and type of fuel
loaded into a cask is monitored and recorded, the cask is closed, and
a tamper-proof seal is applied. As long as the seal is intact and the
records are available, the resulting CoK allows to infer with a great
deal of certainty the contents of the cask. However, even during
routine operations it is conceivable that records are inaccurate or
lost or that seals are compromised. Several methods based on on
neutron or gamma ray detection are under development to restore CoK in
this case, see for instance~\cite{Ziock}.

Here, we envision instead the deployment of an antineutrino
detector, with a fiducial target mass\footnote{The fiducial detector
  mass is the effective mass, after accounting for fiducial volume cuts
and efficiency factors introduced in event reconstruction and analysis.}
of order $\sim 20$~tons, close to the
storage casks for several months.  Using as an example the storage facility at the
Surry Nuclear Power Station in the U.S., where
casks hold 9--16 metric tons of uranium (MTU), we assume
that 50\% of the radioactive material from two of the 15~MTU casks (colored
in red in \cref{fig:surry}) goes missing.  This roughly corresponds to
removing 3\% of the total amount of nuclear waste stored at Surry. We
make no claim that an actual diversion case would have any similarity
to this scenario nor that this could occur as part of routine
operations, it merely serves to indicate the general level of sensitivity
we might expect from antineutrino monitoring.

To determine what it takes to discover such an anomaly, we simulate
the expected number of detected antineutrino events as a function of the
detector position for the two hypothesis ``all storage casks full''
($F$) and ``50\% of nuclear material missing in two casks'' ($M$).  We
use the antineutrino spectrum given by the blue dashed curve in
\cref{fig:spectra} (10~years after discharge) and the inverse beta
decay cross sections from \cite{Vogel:1999zy}. Neutrino oscillation
effects, though small, are taken into account, with the oscillation
parameters given in \cite{Gonzalez-Garcia:2014bfa}. The rate of
antineutrino events per ton of fiducial detector mass and per MTU of
source mass is
\begin{align}
  N_\nu = 5.17\;\text{yr}^{-1} \, \text{ton}^{-1} \, \text{MTU}^{-1}
            \times \bigg( \frac{\text{10\,\text{m}}}{d} \bigg)^2 \,,
\end{align}
where $d$ is the distance between the source and the detector (both treated
as point-like).  This number
depends mildly on the time after discharge and is for instance reduced by $\sim
5\%$ one year later. In the following, we will 
always assume measurement campaigns lasting one year or less and therefore neglect
this small effect.

The irreducible background to the measurement includes antineutrinos from
running nuclear reactors with an expected event rate of
\begin{align}
  N_\text{bg} = 359\;\text{yr}^{-1} \, \text{ton}^{-1} \, \text{GWth}^{-1}
                \times \bigg( \frac{\text{\text{km}}}{d} \bigg)^2 \,.
  \label{eq:}
\end{align}
For the 5.2~GWth (thermal power) reactor in Surry, located $d \sim 1$~km away
from the envisioned 20~ton detector, this leads to $\sim 37\,300$ antineutrino
events per year.  We take this background into account in our simulations.
Backgrounds from other power stations and from radioactive decays in the Earth
(geo-neutrinos) are smaller by at least a factor $\sim 10^{-4}$,  and we
therefore neglect them.

The dominant reducible backgrounds arise from radioactive decays and cosmic ray
interactions mimicking an antineutrino signal.  With current single-volume
liquid scintillator detectors like Double Chooz, RENO, and Daya Bay, when
deployed at the surface, these backgrounds would be a factor 10--10\,000 larger
than the anticipated antineutrino signal.  Current detectors identify signal
candidates by looking for a delayed coincidence between a primary particle and
a delayed neutron capture. However, they are not able to exploit the spatial
correlations between the primary and delayed signals, nor can they tell whether
the primary particles is a positron, as in inverse beta decay, or a photon or
electron, as in most background events.  Fortunately, these shortcomings could
be overcome in a detector with sufficient spatial resolution to tag positrons
by resolving the two 511\,keV x-rays from their
annihilation~\cite{Safdi:2014hwa}.  Prototypes of detectors with this
capability exist and have been successfully operated in particular by the SoLid
and CHANDLER collaborations \cite{Bowden:2016ntq}. Currently, an improvement of
the signal-to-background ratio by a factor of 1\,000 is achievable, and further
improvements appear feasible with improved shielding and an increased
concentration of neutron capture targets like lithium-6.  It thus appears
plausible that within a few years even the low rate of antineutrinos from
nuclear waste will become detectable in surface detectors.  In the following, we
will assume that this has been achieved by the time the proposed measurements
are carried out, and we will neglect reducible backgrounds.

Events are divided into 0.2\,MeV wide energy bins.  Denoting the number
of signal events expected under the two alternative hypotheses by
$F_i$ and $M_i$, and the number of background events by $B_i$, we
define the test statistic
\begin{align}
  \hspace{-0.8cm}
  \chi^2 \equiv 2 \sum_i \left\{ F_i - M_i
                  + (M_i+B_i)\,\log\left[\frac{M_i+B_i}{F_i+B_i}\right]\right\},
  \label{eq:chi2}
\end{align}
which follows a $\chi^2$ distribution.

The results of the analysis are represented by the contours in
\cref{fig:surry} which indicate where the antineutrino detector should be
placed in order to establish the flux deficit at the 90\% confidence
level with 20\,ton\,yrs, 40\,ton\,yrs, and 80\,ton\,yrs of exposure,
respectively.  We see that the detector needs to
be placed within $\sim 50$~meters of the affected casks in order to
collect the $\sim 4000$ events needed for the measurement.

\section{Application to long-term storage facilities}

Above-ground
storage of spent nuclear fuel, while widely used, is only a temporary
solution, and the long-term goal must be to establish underground
repositories that can keep radioactive material out of the biosphere
for $10^4$--$10^6$~years~\cite{NRC:1995}.  The usefulness of
antineutrino detectors at such geological repositories is limited by
the low antineutrino fluxes after strontium-90 has decayed away (half-life
28.8\,yrs).  Moreover, in order not to disturb the repository,
construction of antineutrino detectors seems feasible and useful only
at distances of order 100~meters or larger.

\begin{figure}
  \begin{center}
    \includegraphics[width=0.8\columnwidth]{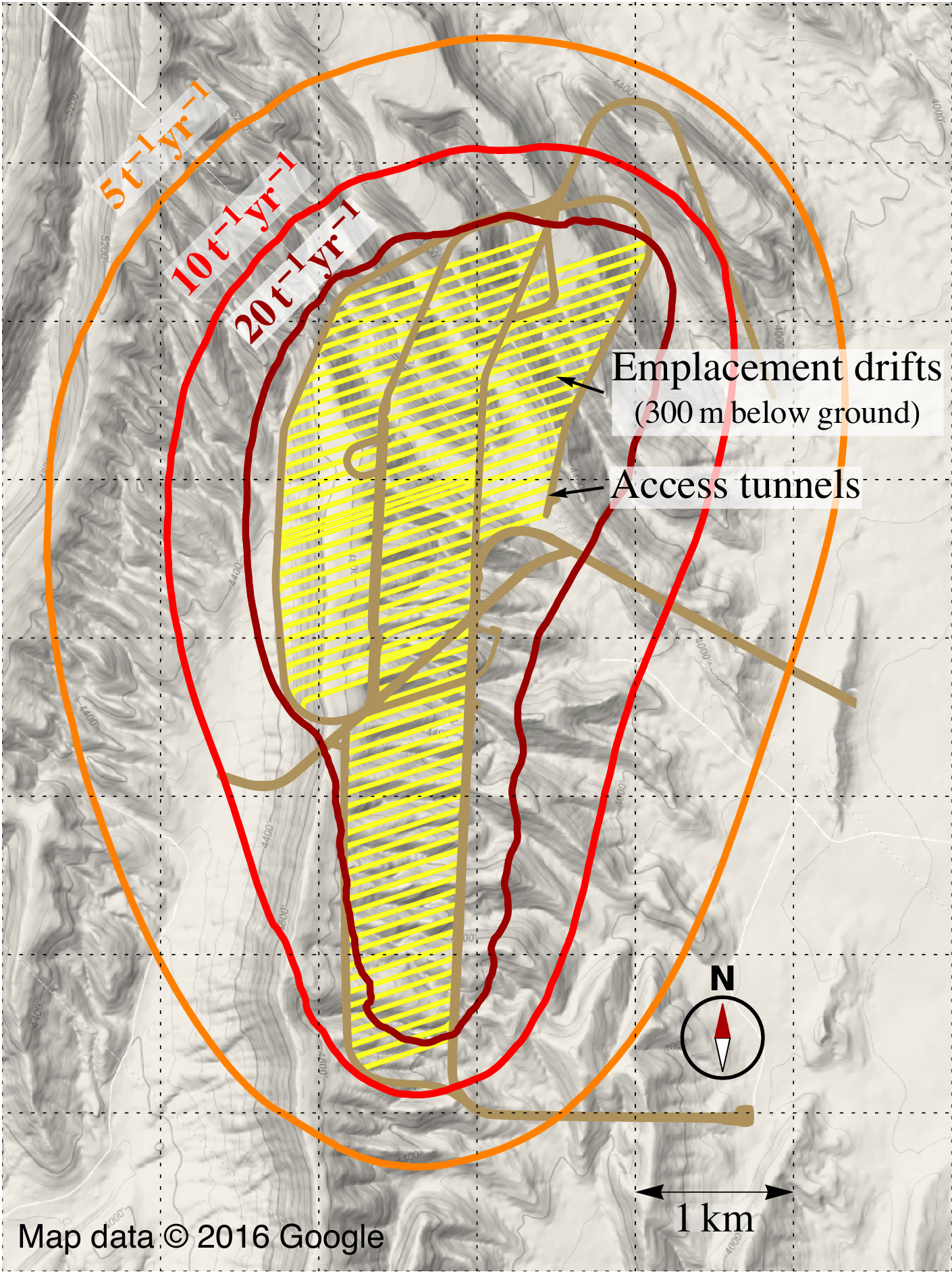}
  \end{center}
  \caption{The planned long term storage facility at Yucca mountain.
    The yellow grid indicates the drifts holding the radioactive material at a
    depth of 300\,m below the surface, while red and orange contours
    show the expected antineutrino count rates for a detector at the surface.}
  \label{fig:yucca}
\end{figure}

To illustrate the prospects of detecting antineutrinos from a
geological nuclear waste repository, we show in \cref{fig:yucca} the
signal event rates expected at the proposed Yucca mountain
repository in Nevada, which would hold 70\,000\,MTU of radioactive
material, stored $\sim 300$\,m underground.  We see that even a small
detector ($\sim 10\,\text{tons}$) located at the surface would see an
appreciable event rate. With a kiloton-scale instrument like KamLAND
\cite{Alivisatos:1998it} or the planned JUNO experiment~\cite{Djurcic:2015vqa},
count rates would be significantly larger, especially
when such a detector is placed in an underground location closer to the
repository.  Even then, however, it would only be possible to detect
cataclysmic disruptions of the repository.  More typical (but
nevertheless highly dangerous) failure scenarios that involve the
leakage of only a small amount of nuclear material into the
surrounding soil would not be detectable. This may change, however,
once detector technology with better directional sensitivity becomes
available (see below).

\section{Leakage of nuclear material at the Hanford site}

Sometimes,
nuclear oversight agencies are faced with the challenge to secure or
decommission a nuclear waste repository in which the contents, and perhaps even
the underground location, of storage casks are not known.  An example is the
Hanford site in the state of Washington (USA), where plutonium for military
purposes was produced from 1944 to 1987.  At Hanford, a major problem is the
leakage of storage containers for high level nuclear waste, leading
to radioactive contamination of ground water~\cite{Fuller:2005,Rockhold:2012}.

Consider first a scenario where the location of storage tanks is
known, but their precise content is not.  We will focus on one
particular array of storage tanks at Hanford, the T~tank
farm~\cite{Fuller:2005}, which consists of 16 tanks, arranged in a $4
\times 4$ grid measuring $\sim 120\,\text{m} \times 80\,\text{m}$ and
originally containing between 0.2 and 5\,MTU of spent fuel each.
We assume the nuclear material in the tanks was discharged from a
reactor 50~years ago. With a detector
placed 30~meters from the most massive (5\,MTU) tank, and taking into
account background antineutrinos from other storage tanks and from the Columbia
nuclear power plant (30\,km away, 3.5\,GW thermal power), the amount of
material in that tank can be measured with an uncertainty of $\pm
2.1\,\text{MTU}$ for an exposure of 20\,ton\,yrs ($35$~signal events)
and with an uncertainty
of $\pm 1.1\,\text{MTU}$ for an exposure of 80\,ton\,yrs ($140$~signal events).
The age of
the nuclear material (i.e.\ the time after discharge) can be
determined to lie between 44\,yrs and 54\,yrs with 80\,ton\,yrs of exposure,
assuming the true age is 50\,yrs.

Assume now that a fraction of the radioactive material in the most massive tank
is slowly leaking out.  We model this situation by reducing the inventory of
the tank and introducing a secondary point source containing the leaked
material 20~meters below its original location.  Detecting such leakage seems
unfeasible with established detector technologies, but requires antineutrino
detectors that not only measure energy, but also the direction of incoming
antineutrinos. Some preliminary efforts in this direction have
been undertaken~\cite{Tanaka:2014, Safdi:2014hwa, Li:2016yey}, but a working
detector is still far off.  As one of the goals of the present study is to
motivate further R\&D in this field, we will in the following
assume the availability of a compact detector with an expected
angular resolution down to $\mathcal{O}(10)$~degrees~\cite{Safdi:2014hwa,Brdar:2017vxy}.
We bin events in the
cosine of the zenith angle, $\cos\theta$, (5~bins) and the azimuth angle
$\phi$ (9~bins).  We use a highly simplified model of angular smearing
in terms of a Gaussian with a width of 20~degrees. For
simplicity, we integrate over energy, assuming that the time of
discharge and thus the antineutrino energy spectrum are known already.  We
estimate that by deploying a directionally sensitive 20\,ton detector
at a distance of 30~meters from the damaged tank, leakage of 55\% of
the tank's content can be discovered at 90\% CL after $12$ months
($30$~signal events).
With an 80\,ton detector, detection of 25\% leakage is possible.

\section{Radioactive spill at Hanford building~324}

As a further application scenario, we consider an actual spill of radioactive
material that happened in October 1986 in a radiochemical plant at Hanford
known as Building~324.  At the time, a large amount of strontium-90 and
caesium-137, with a total activity of 1.3\,MCi was released from a hot cell.
Half of it leaked into the ground and is now presumed to be located within a
$10\,\text{meter} \times 10\,\text{meter}$ area about 2~meters below ground
level~\cite{Rockhold:2012}.  In 2010, a pit was excavated $\sim 20$~meters from
the spill to drive long steel pipes into the affected area, thus allowing the
deployment of temperature and activity sensors.  This method has the
disadvantage that it allows moisture to enter the contaminated soil, which may
ultimately allow radioactive material to seep further into the ground, possibly
reaching ground water levels.  For future incidents of this type, we therefore
consider the deployment of antineutrino detectors for remote sensing.  Modeling
the spill as a point source at a depth of 2\,m, and assuming the availability of an
80\,ton antineutrino detector with angular sensitivity located 30\,m away at the
same depth, we find that a further downward shift of the nuclear material by
3.5\,m is detectable at the 90\% CL after one year of exposure.

\section{Localizing nuclear waste containers}

\begin{figure}
  \centering
  \begin{tabular}{c}
    \includegraphics[width=.7\columnwidth]{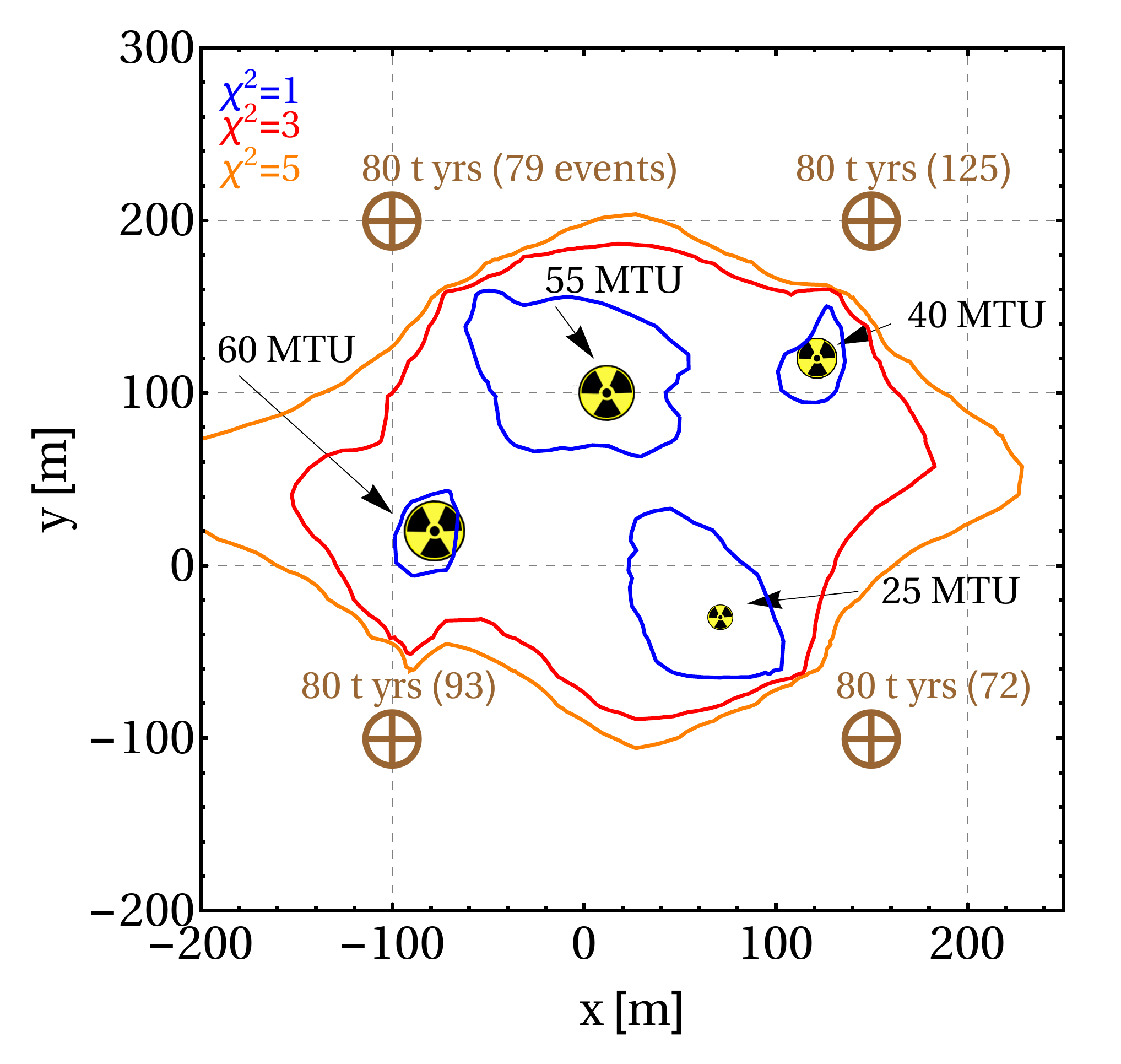} 
  \end{tabular}
  \caption{Using antineutrino detectors (brown $\oplus$ symbols) to localize nuclear
    waste storage casks (radioactive hazard symbols).
    Colored contours indicate the accuracy with which
    casks can be localized (see text for details).
    We have assumed an exposure of $80\,\text{t}\,\text{yrs}$ per detector,
 and we have used the antineutrino spectrum expected 50~years after discharge from
    a reactor.
  }
  \label{fig:hanford}
\end{figure}

Let us now turn to a more
speculative scenario where neither the exact location nor the contents
of storage casks are known. This could happen, for instance, when
documentation is lost and localization using other
methods like ground-penetrating radar is not feasible, for instance in
a scenario where many casks are buried underground, but only few
contain high level nuclear waste.  We envision successive or
simultaneous deployment of 80\,ton antineutrino detectors on a
two-dimensional grid with a spacing of 250\,m.  We again assume
angular sensitivity, but since the distance between detectors and
sources is large, the zenith angle measurement is irrelevant and can be
discarded.  \Cref{fig:hanford}
illustrates the outcome of such an analysis for four randomly
placed storage casks of unknown content and for an exposure of
80\,ton\,yrs per detector.
Using a stochastic optimization method, we fit the positions
$(x_j, y_j)$ and activities $m_j$ of the four sources.  Colored contours
show the dependence of the test statistic $\chi^2$
(defined in analogy to \cref{eq:chi2}) on the fit values of $(x_1, y_1)$,
with the other $(x_j, y_j)$ as well as all $m_j$ allowed to float.
We see that storage casks can be localized to within tens of meters.  For
further refinement of their position (for instance in order to guide
cleanup efforts), the procedure can be repeated with
detectors moved closer to the source positions determined in the initial scan.
With a $\sim 30$\,m spacing between detectors, the position
of each source can be determined to $\mathcal{O}(\text{m})$ accuracy.

\section{Summary}

We have calculated the antineutrino flux and spectrum from
spent nuclear fuel, and have used these results to outline possible
applications of antineutrino detectors in monitoring and managing nuclear waste
repositories.  We have shown that, in a specific diversion scenario at a
dry cask storage facility, installation of an antineutrino
detector could allow oversight agencies to remotely detect leakage or theft
of the stored nuclear waste. Further study is needed to assess the applicability
of the method to other diversion scenarios.
At long term geological repositories, a
significant antineutrino flux is expected, but detecting realistic anomalies such
as leakage of a small amount of radioactive material would require
advanced detector technologies with angular sensitivity.
Such detectors could also help in the decommissioning of nuclear
installations like the Hanford site, where they would allow for the
localization of nuclear material and for the characterization of spills.

\section*{Acknowledgments}

PH thanks Fermilab for hospitality during the
completion of this manuscript. This work was in part supported by the U.S.
Department of Energy under contracts DE-SC0013632 and DE-SC0009973.  The work
of VB and JK is supported by the German Research Foundation (DFG) under Grant
Nos.\ \mbox{KO~4820/1--1} and FOR~2239 and by the European Research Council
(ERC) under the European Union's Horizon 2020 research and innovation programme
(grant agreement No.\ 637506, ``$\nu$Directions'').  VB is also supported by
the DFG Graduate School Symmetry Breaking in Fundamental Interactions (GRK
1581).


\bibliography{nuclear-waste}

\end{document}